\documentclass[reprint,twocolumn, 
            amsmath,amssymb,
            tightenlines,citeautoscript,
            flushbottom, balancelastpage,
            superscriptaddress, floatfix,
            ]{revtex4-1}
\usepackage[utf8]{inputenc}
\usepackage[english]{babel}
\usepackage{hyperref}
\usepackage{etoolbox} 
\usepackage{multibib}
\usepackage{bm}
\usepackage{hyperref}
\usepackage[dvipsnames]{xcolor}
\usepackage[capitalise]{cleveref}
\usepackage{graphicx}
\usepackage{bold-extra}
\usepackage{tabularx}
\usepackage{float}
\usepackage{MnSymbol}
\usepackage{booktabs} 
\usepackage{siunitx}
\usepackage{caption}
\usepackage[version=4]{mhchem}
\usepackage{xfrac}
\usepackage[
    protrusion=true,
    activate={true,nocompatibility},
    final,
    tracking=true,
    kerning=true,
    spacing=true,
    factor=1100]{microtype}

\definecolor{gregRed}{RGB}{209, 0, 11}

\newcommand{\gregsout}{\bgroup\markoverwith{\textcolor{gregRed}{\rule[0.5ex]{2pt}{0.4pt}}}\ULon}

\hypersetup{
 pdfauthor={Gregory Moille},     % author
 pdfcreator={Gregory Moille},   % creator of the document
 pdfproducer={LaTeX}, % producer of the document
 pdfkeywords={Version 1.0},
 colorlinks=true,       % false: boxed links; true: colored links
 linkcolor=blue,          % color of internal links (change box color with
 citecolor=blue,        % color of links to bibliography
 filecolor=blue,      % color of file links
 urlcolor=blue}

\usepackage{titlesec}
\titleformat*{\section}{\bfseries\Large}
\titleformat*{\subsection}{\bfseries}

\begin{document}

%  _____ ___ _____ _     _____
% |_   _|_ _|_   _| |   | ____|
%   | |  | |  | | | |   |  _|
%   | |  | |  | | | |___| |___
%   |_| |___| |_| |_____|_____|
%
% \title{Arbitrary Microring Dispersion Engineering for Ultrabroad Frequency Combs: Photonic Crystal Microring Design Based on Fourier Synthesis}
\title{Fourier Synthesis Dispersion Engineering of Photonic Crystal Microrings for Broadband Frequency Combs}%Concept and Limit of Fourier Synthesis Dispersion Engineering of Photonic Crystal Microring Frequency Combs}

%     _   _   _ _____ _   _  ___  ____  ____
%    / \ | | | |_   _| | | |/ _ \|  _ \/ ___|
%   / _ \| | | | | | | |_| | | | | |_) \___ \
%  / ___ \ |_| | | | |  _  | |_| |  _ < ___) |
% /_/   \_\___/  |_| |_| |_|\___/|_| \_\____/
%
\author{Gr\'egory Moille}
\email{gmoille@umd.edu}
\affiliation{Joint Quantum Institute, NIST/University of Maryland, College Park, USA}
\affiliation{Microsystems and Nanotechnology Division, National Institute of Standards and Technology, Gaithersburg, USA}
\author{Xiyuan Lu}
\affiliation{Joint Quantum Institute, NIST/University of Maryland, College Park, USA}
\affiliation{Microsystems and Nanotechnology Division, National Institute of Standards and Technology, Gaithersburg, USA}
\author{Jordan Stone}
\affiliation{Joint Quantum Institute, NIST/University of Maryland, College Park, USA}
\affiliation{Microsystems and Nanotechnology Division, National Institute of Standards and Technology, Gaithersburg, USA}
\author{Daron Westly}
\affiliation{Microsystems and Nanotechnology Division, National Institute of Standards and Technology, Gaithersburg, USA}
\author{Kartik Srinivasan}
\affiliation{Joint Quantum Institute, NIST/University of Maryland, College Park, USA}
\affiliation{Microsystems and Nanotechnology Division, National Institute of Standards and Technology, Gaithersburg, USA}
\date{\today}

\maketitle

% ooo        ooooo            o8o              
% `88.       .888'            `"'              
%  888b     d'888   .oooo.   oooo  ooo. .oo.   
%  8 Y88. .P  888  `P  )88b  `888  `888P"Y88b  
%  8  `888'   888   .oP"888   888   888   888  
%  8    Y     888  d8(  888   888   888   888  
% o8o        o888o `Y888""8o o888o o888o o888o 

\noindent\textbf{\large Abstract}

\noindent Dispersion engineering of microring resonators is crucial for optical frequency comb applications, to achieve targeted bandwidths and powers of individual comb teeth.  However, conventional microrings only present two geometric degrees of freedom -- width and thickness -- which limits the degree to which dispersion can be controlled. We present a technique where we tune individual resonance frequencies for arbitrary dispersion tailoring. Using a photonic crystal microring resonator that induces coupling to both directions of propagation within the ring, we investigate an intuitive design based on Fourier synthesis. Here, the desired photonic crystal spatial profile is obtained through a Fourier relationship with the targeted modal frequency shifts, where each modal shift is determined based on the corresponding effective index modulation of the ring. Experimentally, we demonstrate several distinct dispersion profiles over dozens of modes in transverse magnetic polarization. In contrast, we find that the transverse electric polarization requires a more advanced model that accounts for the discontinuity of the field at the modulated interface. Finally, we present simulations showing arbitrary frequency comb spectral envelope tailoring using our Frequency synthesis approach.

%Dispersion engineering of microring resonator is crucial to achieving the targeted bandwidth with optical frequency comb and the needed power of individual comb teeth for stabilization. Yet microring resonators only present two degrees of freedom -- width and height -- limiting dispersion engineering. We present a new technique where we tune individual resonance frequencies for arbitrary dispersion tailoring. Using a photonic crystal microring resonator coupling both directions of propagation, we investigate intuitive design based on Fourier synthesis. We show that using the refractive index modulation rather than the ring width modulation, one could obtain the photonic crystal profile through a Fourier relationship with the targeted modal frequency shift. We demonstrate this design over dozens of modes in transverse magnetic polarization. In contrast, the transverse electric would require a more advanced model accounting for the discontinuity of the field at the modulated interface. We show arbitrary frequency comb spectral envelope tailoring through simulation using our Frequency synthesis approach. 

\vspace{1ex}\noindent\textbf{\large Introduction}

\noindent Frequency combs based on integrated nonlinear microresonators are a powerful tool to bring metrology outside the lab. They allow for low power consumption and portabilty~\cite{SternNature2018} while maintaining metrological quality~\cite{SpencerNature2018} while in the dissipative Kerr soliton (DKS) regime. Although octave-spanning frequency combs -- needed for carrier-envelope stabilization through self-referencing of the comb -- have been demonstrated~\cite{LiOptica2017,PfeifferOpticaOPTICA2017}, reaching beyond an octave is particularly interesting so that the strong pump can be doubled in self-referencing schemes. Yet, it is extremely challenging, especially at short wavelengths towards the visible. Materials that are used for microcomb generation, including \ce{Si3N4}~\cite{BraschScience2016, NazemosadatPhys.Rev.A2021, YeLaserPhotonicsRev.2022, MoilleNat.Commun.2021,ShenNature2020}, \ce{AlN}~\cite{BruchNat.Photonics2021}, and \ce{LiNbO3}~\cite{WangNatCommun2019}, present increasingly large normal dispersion the shorter the wavelength is~\cite{LeeNatCommun2017, MoilleOpt.Lett.OL2021, MoilleCLEO20222022}. Modal confinement of the light in resonators with wavelength-scale cross-sections adds a geometrical component to the dispersion, which in many cases is enough to compensate for the normal material dispersion. However, a simple rectangular cross-section microresonator does not offer enough degrees of freedom to achieve broad enough anomalous dispersion (needed for bright DKS states) to tackle goals such as spectral bandwidths well-beyond an octave while extending well into the visible. Alternative approaches have been proposed. Among them, multi-pumped DKS~\cite{MoilleNat.Commun.2021} and pulsed-pump resonators~\cite{AndersonNatCommun2022,AndersonOptica2021} have been successful in realizing spectral bandwidths beyond that of conventional DKS microcombs. Yet, these solutions often increase the complexity of the setup required for field deployment of these microcombs. It is thus necessary to create methods through which one can engineer the dispersion of a microring resonator. Approaches with multi-layer material stacks~\cite{DorcheOpt.ExpressOE2020}, complex ring cross-sections~\cite{MoilleOpt.Lett.OL2018} and concentrical rings~\cite{KimNat.Commun.2017} -- each relying on avoided-mode crossings -- have generated much more complex dispersion profiles. However, fabrication has been challenging and may be incompatible with top-down foundry-like mass fabrication processes~\cite{LiuNatCommun2021a}. In addition, broad bandwidth microcombs based on these approaches have not yet been demonstrated. Nevertheless, the concept of an avoided-mode crossing, which relies on mode-coupling, can be harnessed in different fashions, for example, by coupling the clockwise (CW) and counterclockwise (CCW) directions of the same transverse optical mode. Such CW/CCW coupling has been demonstrated by Lu \textit{et al.}, where modulation of the microresonator sidewall creates a photonic crystal that frequency splits a targeted mode without impacting the nearest neighbors, determined by the number of photonic crystal periods within the ring circumference~\cite{LuAppl.Phys.Lett.2014}. Interestingly, nonlinear states such as optical parametric oscillation~\cite{BlackOpticaOPTICA2022, LuOpt.Lett.OL2022, StonearXiv2022} and DKSs can be created in this system~\cite{YuNat.Photonics2021}. Moreover, this photonic crystal ring concept has been expanded to modulation amplitudes far beyond a simple perturbation, where a full band gap of hundreds of gigahertz to several terahertz is resolved, impacting the band structure among several neighboring modes~\cite{LuNat.Photon.2022}. More importantly for this work, it has been shown that it is possible to introduce multi-period photonic crystal patternings that create controlled frequency splittings for a few (up to 5) targeted modes~\cite{LuPhotonicsRes.2020}. In particular, sinusoidal ring width modulations corresponding to multiple single modes targeted simultaneously (\textit{i.e.} with different amplitudes and modulation periods) can be summed with limited impact to the other modes. In summary, ref.~\cite{LuPhotonicsRes.2020} lays out a methodology based on Fourier synthesis for microring dispersion engineering~\cite{MoilleConf.LasersElectro-Opt.2022Pap.STh2F32022}, where a spectral profile of mode coupling and corresponding shift of several individual modal frequencies follows a discrete Fourier transform of the ring width modulation. 

\begin{figure*}[t]
    \centering
    \includegraphics[width=\textwidth]{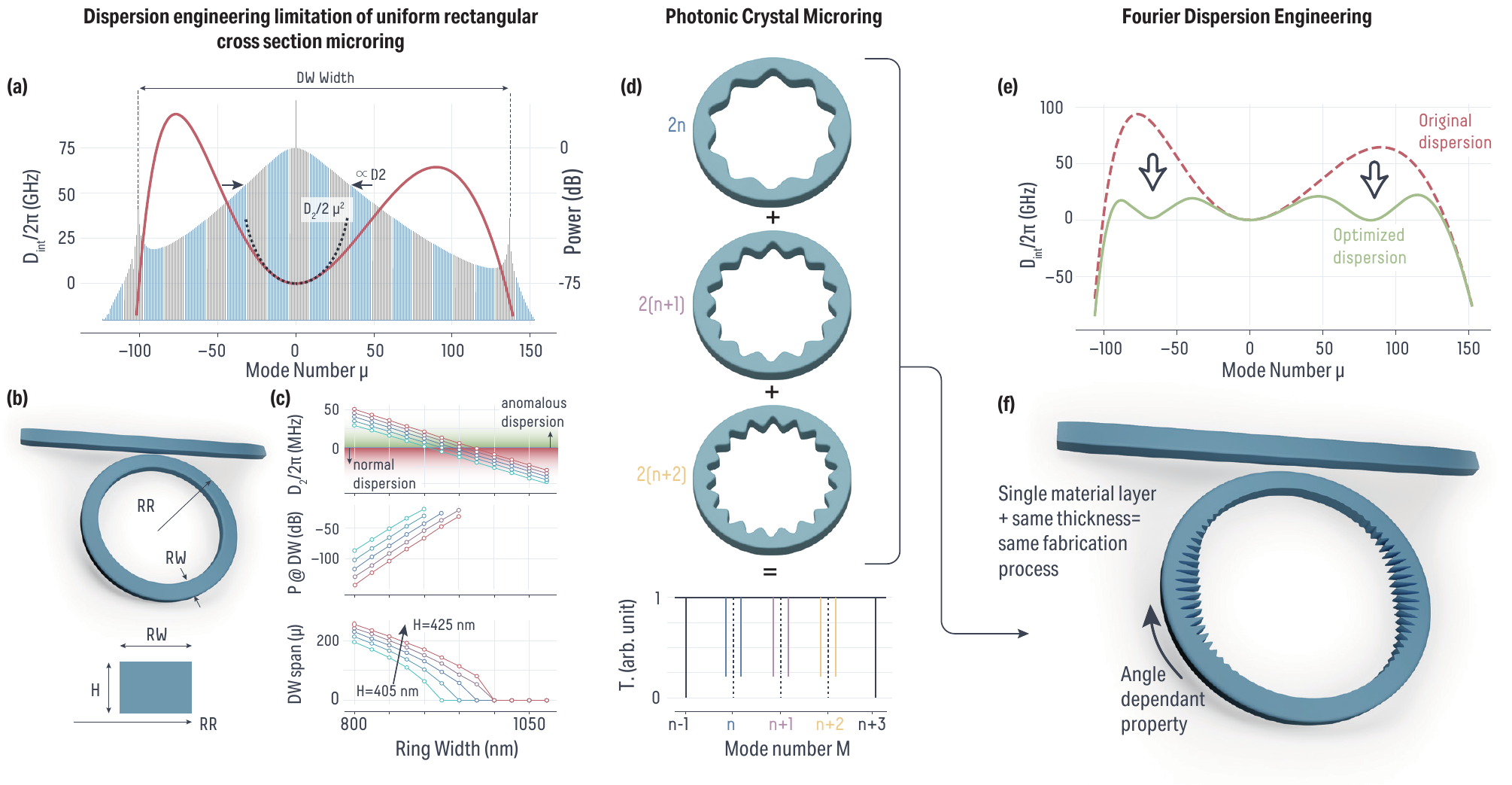}
    \caption{\label{fig:1} %
    \textbf{Photonic crystal patterning for broadband microring dispersion engineering. (a)} Simulated frequency comb (blue) assuming the $D_\mathrm{int}$ profile in red similar to that in ref.~\cite{LiOptica2017}, and suitable for octave span thanks to dual-DWs. Positive quadratic dispersion ($D_{2}>0$) results in the $\mathrm{sech}^2$ comb envelope near the pump, while further away, higher-order dispersion allows for zero-crossings where $D_{int}=0$. This results in comb teeth on resonance, locally enhancing the power through DWs -- and driving the overall bandwidth of the comb, while $D_2$ dictates the width of the $\mathrm{sech}^2$ envelope. The soliton power (following the $\mathrm{sech}^2$ envelope) at $D_\mathrm{int}=0$ impacts the generated DW power. %
    \textbf{(b)} Standard microring resonator where three parameters are available for dispersion engineering, ring radius ($RR$), ring width ($RW$) and material thickness ($H$). %
    \textbf{(c)} Variation of $D_2$, power in the DW, and separation between the two potential DW locations (in units of mode number) as a function of ring width. These plots illustrate the trade-off between potential microcomb bandwidth and DW power, where the largest comb bandwidth (bottom panel) requires large $D_2$ (top panel) and hence a sharper $\mathrm{sech}^2$ envelope that leads to lower available power at the $D_\mathrm{int}$ zero crossing for DW enhancement (middle panel). %
    \textbf{(d)} Concept of a photonic crystal ring where a sinusoidal modulation of the ring at a period $2n$ will split the mode $M=n$ through CW/CCW coupling, with a strength dependent on the modulation amplitude. Within a certain regime of modulation amplitude, other azimuthal modes are not disturbed. Therefore, one can simply add modulations with different periods and amplitudes to selectively target different modes $M$ with desired frequency splittings ($T$= transmission in arbitrary units). %
    \textbf{(e)} The frequency splittings modify $D_\mathrm{int}$, resulting in a higher $D_\mathrm{int}$ branch and a lower $D_\mathrm{int}$ branch (shown in green). When a sufficiently large number of modes are split, the lower $D_\mathrm{int}$ branch can be flattened (i.e., largely reducing the maximum $D_\mathrm{int}$ value) to bypass the $D_2$ vs. potential comb bandwidth trade-off. %
    \textbf{(f)} The resulting ring resonator has a ring width modulation profile that is the inverse DFT of the splitting of each mode that one wants to design, essentially equivalent to a Fourier synthesis approach. As a result, many split modes will produce a very localized ring-width modulation. This introduces challenges that we discuss throughout the rest of the paper.}
\end{figure*}

Yet, whether a vast number of mode-couplings through Fourier synthesis -- which remains a perturbative appraoch -- can be implemented in a predictive manner that yields precise and efficient dispersion engineering consistent with the needs of, for example, broadband microresonator frequency combs is still unclear. In this work, we propose to answer this question with an in-depth study of photonic-crystal-mediated microring dispersion engineering in the limit of large (tens of gigahertz) spectral shifts. We demonstrate that the previously utilized simple analysis that links the modal frequency shifts directly to ring width modulation is inaccurate in our regime of interest. In contrast, modulation of the ring effective refractive index, which is mapped (nonlinearly) to a ring width modulation, is a better approach. We also show that the polarization considered greatly impacts the validity of the perturbative approach with which this mode-by-mode dispersion engineering is predicted using our Fourier synthesis model. This limit arises from the boundary conditions on the dominant electric field components at discontinuous boundaries. Consequently, the transverse magnetic polarization is more suited for predictive dispersion engineering using our straightforward Fourier synthesis approach, while the transverse electric polarization may require an approach based on full three-dimensional numerical simulations of Maxwell's equations in modulated microring structures. Using our technique, we fabricate silicon nitride photonic crystal microrings in which dozens of resonances are shifted in a controlled fashion by up to 50~\unit{GHz}, compatible with the integrated dispersion mitigation needed for broadband (e.g., octave-spanning) combs, and in good agreement with simulations. Finally, we use coupled Lugiato-Lefever equation modeling to predict the spectral behavior of microcombs that can be generated using our dispersion technique, and in particular, the possibility of considerably extending their bandwidth and the possibility of creating multi-color Bragg solitons.

\noindent\vspace{1ex}\textbf{\large Results}

\noindent\textbf{Towards a broad and flat microcomb spectrum: motivating photonic crystal dispersion engineering}. Under the right conditions of pump power and detuning, microcombs based on the third-order optical nonlinearity can support DKS states~\cite{KippenbergScience2018}. The usual way of studying the dynamics of such a system is through the Lugiato-Lefever equation~\cite{LugiatoPhys.Rev.Lett.1987} which is essentially a dissipative nonlinear Schr\"odinger equation~\cite{HanssonPhys.Rev.A2013} that takes into account the microring resonator's periodic boundary conditions while operating under a slow varying envelope (mean-field) approximation~\cite{ChemboPhys.Rev.A2013}. The single DKS solution in the anomalous dispersion regime follows the well-known hyperbolic secant (sech) spectral envelope (sech\textsuperscript{2} for spectral intensity). To quantify the resonator dispersion, the community usually opts for a Taylor expansion and defines a quantity termed the integrated dispersion $D_\mathrm{int} = \sum_{k>1}\frac{D_k}{k!} \mu^k = \omega_\mathrm{res} - (\omega_0 + D_1\mu)$, with $D_1$ being the linear repetition rate at the pumped mode with resonant frequency $\omega_0$, and $\mu$ the azimuthal mode number referenced to the pumped mode. The higher order dispersion terms $D_k$ are of great importance as they drive the shape of the integrated dispersion and, ultimately, the properties of the microcomb. The sech\textsuperscript{2} comb envelope width is inversely proportional to $D_2$, which must be positive for anomalous dispersion. The odd terms $D_{2k+1}$ drive the recoil of the soliton, resulting in the drift of its repetition rate away from the linear one. The even terms $D_{2k}$ are responsible for symmetric zero crossings of $D_\mathrm{int}$ and yield dual dispersive waves (DWs) as comb teeth become resonant at these modal frequencies. The experimental demonstrations of DWs~\cite{BraschScience2016} have fundamentally changed the landscape of microcombs by bypassing the sech\textsuperscript{2} width driven by $D_2$ and expanding it to octave span~\cite{LiOptica2017,PfeifferOpticaOPTICA2017}. In this work, we will refer to the frequency span between the $D_\mathrm{int}=0$ frequencies as the `dispersive wave span' of the microcomb [\cref{fig:1}(a)-(b)]. Although these DWs have been the key enabler for broadband microcombs, the power in these modes relies on the available power in the sech\textsuperscript{2} soliton envelope. Therefore, if $D_2$ is too large, leading to a sharp comb envelope close to the pump, the power available at the DW locations will be insignificant for resonant enhancement. This encapsulates the dispersion engineering challenge for broadband integrated frequency combs: getting as broad as possible the $D_\mathrm{int}$ zero-crossings while keeping the $D_2>0$ portion as flat as possible.

Guided-wave photonics results in wavelength-dependent light confinement: the longer the wavelength, the larger the mode and less confined it is. This unique feature allows one to tailor the dispersion beyond the material dispersion. For a microring resonator, there are essentially three user-defined input parameters: ring radius ($RR$), ring width ($RW$), and thickness ($H$) [\cref{fig:1}(b)]. The $RR$ mainly acts on $D_1$ and typically has little influence on the higher order dispersion terms. Therefore, most dispersion engineering efforts focus on $RW$ and $H$. However, with only these two parameters available, the DW span and $D_2$ increase together [\cref{fig:1}(c)]. Although increasing the comb width is the ultimate goal, increasing $D_2$ reduces the power available at the $D_\mathrm{int}$ zero-crossings, ultimately preventing useful DWs from forming. Thus, with current dispersion engineering approaches, an apparent trade-off exists.

%and tread-off needs to be found: either broad comb or small $D_2$.

In recent years, a modified microring resonator has been developed, where modulation of the ring width allows for coherent backscattering between the clockwise (CW) and counter-clockwise (CCW) traveling wave modes~\cite{LuAppl.Phys.Lett.2014,LuPhotonicsRes.2020}. It has been demonstrated, using a perturbative approach theoretically and verified experimentally, that a single harmonic modulation of the ring width results in a single azimuthal mode of a given transverse spatial mode family experiencing this coupling~\cite{LuAppl.Phys.Lett.2014}. These two coupled modes hybridize into symmetric and anti-symmetric modes, creating a mode splitting proportional to the amplitude of the modulation [\cref{fig:1}(d)]. In the unmodulated ring, the necessity to match the field phase after one round trip results in a set of azimuthal modes described by a mode number $M \in \mathbb{Z}$. When the modulation is applied, a new spatial period $\pi RR/M_0\ll 2\pi RR$ becomes important. Here, the modulation consists of $2M_0$ periods around the ring circumference. In the language of photonic crystals (PhCs)~\cite{joannopoulosPhotonicCrystalsMolding2008}, the frequency splitting created by coupling of the CW and CCW azimuthal modes at $\pm M_0$ is a frequency band gap, with the Brillouin zone folded at these points as well. These modulated devices can thus be referred to as photonic crystal (PhC) microrings, and in the limit of strong modulation, large band gaps of several terahertz have been demonstrated while maintaining high optical quality factor ($Q$)~\cite{LuNat.Photon.2022}. In addition, smaller modulation PhC microrings have been used to change the dynamics of DKS formation~\cite{YuNat.Photonics2021}. For smaller modulations, it has also been shown that the splitting is linearly proportional to the modulation amplitude employed, and that the CW/CCW coupling is nearly absent for all modes near the targeted mode $M_0$~\cite{LuPhotonicsRes.2020}.
Therefore multiple mode splittings (PhC modulations) can be implemented on a single ring. Following ref.~\cite{LuPhotonicsRes.2020}, the design rules then become simple: a straightforward sum of individual modulation patterns gives the microring ring width modulation to implement [\cref{fig:1}(d)], such that $RW_\mathrm{mod} = \sum_m A(m) \mathrm{cos}\left(2m(\theta+\varphi)\right)$ with $A(m)$ the modulation intensity of each targeted mode $m$ and $\theta$ the microring azimuthal angle. Therefore, the total ring width modulation in this scheme is, per definition, the discrete inverse Fourier transformation (DFT) of the modal coupling envelope $A(m)$. 

The frequency shift created by the CW/CCW coupling can be used to modify the dispersion of the resonator locally. The integrated dispersion is then shifted by half the resonance splitting at each coupled mode, creating two bands. One band is pushed toward higher $D_\mathrm{int}$, and one reduces the value of $D_\mathrm{int}$. The latter can overcome the aforementioned trade-off in large $D_2$ and large potential comb bandwidth if the Fourier synthesis design approach allows for predictive mode shifts in the tens of gigahertz range to counterbalance the maximum $D_\mathrm{int}$ value of an octave-spanning comb. By carefully engineering the coupling, one could create flatter integrated dispersion while obtaining a large comb width [\cref{fig:1}(e)]. The resulting microring resonator undergoes a local ring width modulation consistent with the inverse DFT [\cref{fig:1}(f)]. Such a system is compatible with standard single device layer fabrication as only the ring width as a function of angle is modified, and no other fabrication steps or materials are added.

\begin{figure*}[t]
    \centering
    \includegraphics[width=\textwidth]{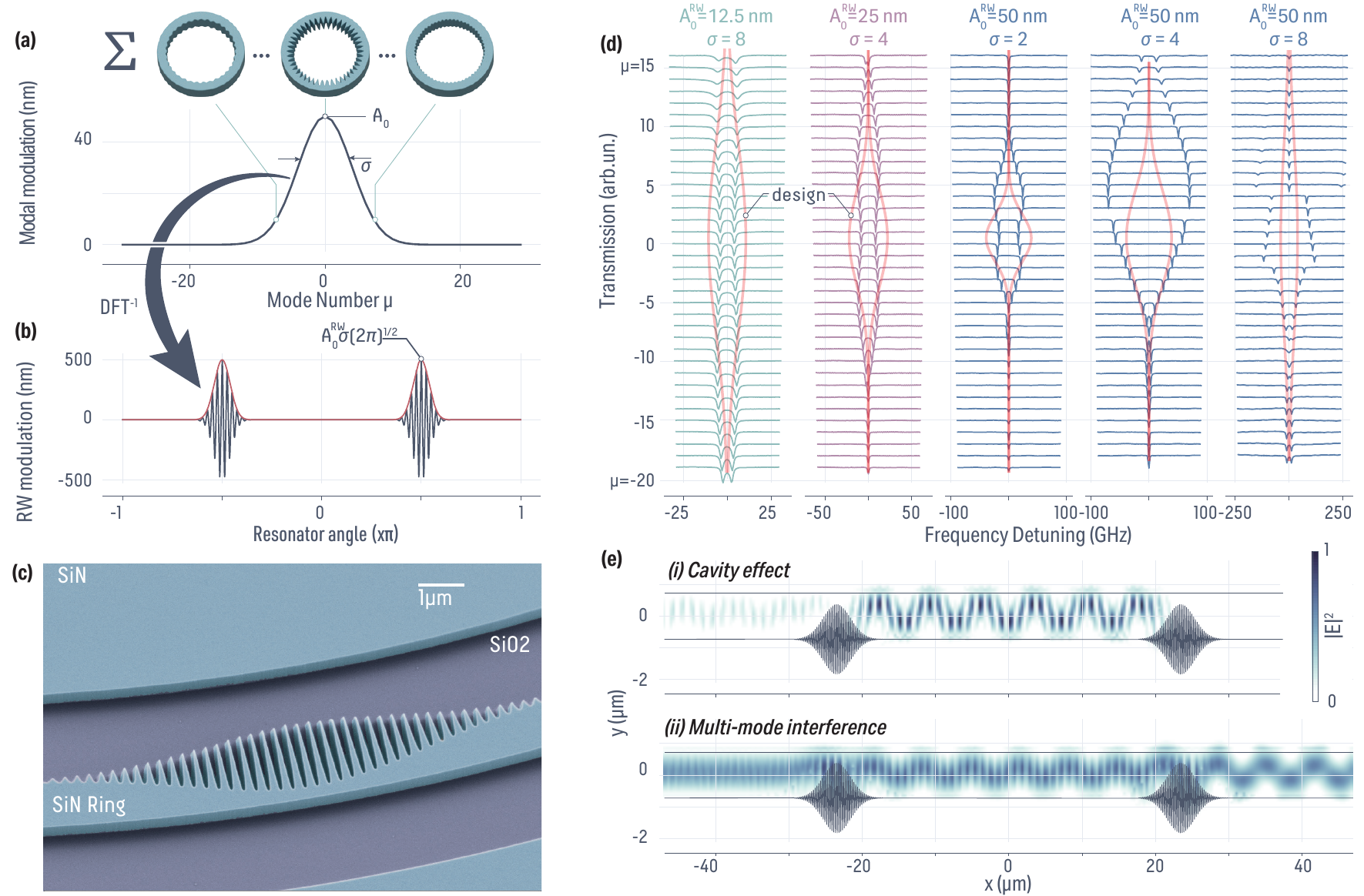}
    \caption{\label{fig:2} %
    \textbf{Multiple mode-splitting from direct Fourier synthesis from ring width modulation. } %
    \textbf{(a)} Modulation amplitude for each targeted mode following a Gaussian of width $\sigma$ and maximum $A_0$. Under the approximation that the $RW$ modulation is linear with the mode splitting, this corresponds to the modal envelope of the CW/CCW coupling. %
    \textbf{(b)} Following the Fourier synthesis of ref~\cite{LuPhotonicsRes.2020,MoilleConf.LasersElectro-Opt.2022Pap.STh2F32022}, the sine modulation of each mode can be summed up, resulting in a discrete inverse Fourier transform for the total ring width modulation with the azimuthal angle. It is worth noting that the maximum $RW$ modulation is far greater than the maximum modal modulation in (a). %
    \textbf{(c)} SEM image of the fabricated Fourier synthesis PhC ring following the Gaussian modal coupling envelope. The resonator is an air-clad \ce{Si3N4} microring of ring radius $RR=60$~\unit{\um}, ring width $RW=1450$~\unit{\nm}, and thickness $H=420$~\unit{\nm} on top of \ce{SiO2}. %
    \textbf{(d)} Experimental measurement of the mode splitting for different widths and amplitudes of the modal Gaussian envelope. The solid line represents the designed CW/CCW coupling, which does not follow the experimental results. \textbf{(e)} Two effects that play a critical role in the discrepancy between design and experiment. (top panel) A cavity effect occurs when the total $RW$ modulation becomes comparable with the nominal $RW$, allowing a local cut-off and creating a Fabry-P\'erot-like cavity. This cavity will couple with the traveling mode, yielding another shift of the resonance. (bottom panel) Apart from this cavity-like resonance, higher-order modes are coupled through the modulation, resulting in avoided mode-crossing(s) that impact the position of the resonances. %
}
\end{figure*}

\vspace{1ex}\noindent\textbf{Limitations of a direct implementation of  the ring width modulation} In this section, we seek to verify the above approach for predictive dispersion engineering. We fabricate devices with $H=440$~\unit{nm} thick silicon nitride (\ce{Si3N4}), nominal ring width $RW=1450$~\unit{nm}, and radius $RR=60$~\unit{um} on top of silicon dioxide (\ce{SiO2}) without any top cladding other than air. We implement a targeted Gaussian modal envelope (in mode number space) defined by $RW_\mathrm{mod}(\mu) = A_0^\mathrm{RW} \mathrm{exp}^{-(\mu/\sqrt{2}\sigma)^2}$, where $\mu = M - M_0$ is the mode number relative to a centered one set at $M_0=392$ [\cref{fig:2}(a)], and $A_0^\mathrm{RW}$ is chosen based on the desired maximum frequency splitting. At the moment, only the fundamental transverse electric mode (TE\textsubscript{0}) will be considered, which we have determined through finite element method (FEM) simulation to be at 193.6~\unit{THz} at $M_0$. In the layout, we vary the maximum amplitude of the Gaussian $A_0$ and its width $\sigma$ while other parameters are fixed. The resulting summed $RW$ modulation profile follows, as expected, the inverse DFT profile (Fig.~\ref{fig:2}(b)). The envelope of the $RW$ modulation is therefore also a Gaussian, and the mapping between frequency splitting of a given mode and real-space RW modulation amplitude is linear, based on the simple perturbative analysis from ref.~\cite{LuPhotonicsRes.2020}. After summing these different modulations, the maximum real-space modulation of the ring width is $\mathrm{max}[RW_\mathrm{mod}(\theta)] = A_0^\mathrm{RW}\sigma \sqrt{2\pi}$, and is much larger than $A_0^\mathrm{RW}$, which is the maximum of the modal Gaussian profile [\cref{fig:2}(a)-(b)], and the consequences of which will be discussed shortly. The maximum $RW$ modulation could be reduced if a phase shift is applied to each modal modulation. Yet, the symmetric and anti-symmetric modes arising from CW/CCW coupling are intrinsically standing waves, with their spatial mode profiles along the azimuthal direction locked by the node and anti-node positions of the modulation. Applying such a phase shift might not be desirable as it could reduce nonlinear interaction between between modes. Thus, we will only study the extreme modulation case where $\varphi = 0$ for Fourier synthesis dispersion engineering which if demonstrated to work should be applicable for arbitrary $\varphi$. A zoom-in scanning electron microscope image of a fabricated device, created using high-resolution electron-beam lithography, is shown in \cref{fig:2}(c). 

We proceed with transmission spectroscopy measurements between 185.18~THz and 198.6~THz (\textit{i.e.} 1510~\unit{nm} and 1620~\unit{nm}) of the ring resonators with different modulation amplitude and width, where each resonance is spaced by a free spectral range of 398~\unit{GHz}. A stacked-up representation of the transmission around each resonance is represented in \cref{fig:2}(d), highlighting the spectral profile of the modal coupling obtained (i.e., mode splitting vs. mode number). However, while overlapping this experimental data with the originally designed coupling from the direct inverse DFT, it is evident that none of the experimental data match. In particular, multiple behaviors are reported with the ``collapsing'' of the splitting around $\mu=0$ for small modulation amplitude $A$ and/or small width $\sigma$, while at large $A$ and large $\sigma$, the modal coupling becomes much more unpredictable. The maximum coupling for each design has been calibrated with a single mode coupling at $\mu = 0$. It is worth reporting that $A_0^\mathrm{RW}=12.5$~nm and $\sigma =8$ yields $\mathrm{max}[RW_\mathrm{mod}(\theta)] = 250$~nm, which is far from the upper range of what has been previously experimentally verified and confirmed to match predictions based on a perturbative approach~\cite{LuPhotonicsRes.2020}. 

We propose that the discrepancy between designs and experiments comes from two fundamental interactions in the PhC ring, which we highlight through finite-difference time-domain (FDTD) simulations. Here, we consider a simple FDTD simulation whose intent is to provide a qualitative assessment of these complicating factors.  To highlight qualitative features while retaining a small enough structure to avoid convergence and/or simulation time issues, we consider a waveguide containing two highly modulated regions, as shown in Fig.~\ref{fig:2}(e). First, under a total modulation where the $RW$ becomes small enough, a cavity-like effect will occur with boundaries localized at the local modulated sections of the ring [\cref{fig:2}(e,i)]. Essentially the propagating modes -- either CW or CCW -- will also be coupled to this cavity mode, making the mode splitting much more complex. We believe such an effect is mostly seen in the higher modulation case, as highlighted in $A_0^\mathrm{RW}=50$~nm and $\sigma =8$. The second effect that comes into play is the multi-mode interference. The $RW$ modulation, in particular reducing the ring width, allows for a non-zero overlap between the higher-order transverse spatial modes of the unperturbed ring. Similar to an unoptimized racetrack resonator~\cite{JiCommunPhys2022}, this yields a coupling between these modes, which adds up to the CW/CCW one and creates a much more complex pattern of mode splitting. This multi-mode interference effect is also associated with higher radiation losses and quality factor asymmetry, as can be seen for $\mu>5$ in the case of $A_0^\mathrm{RW}=50$~nm and $\sigma =8$ in \cref{fig:2}(d)

\begin{figure*}[t]
    \centering 
    \includegraphics[width=\textwidth]{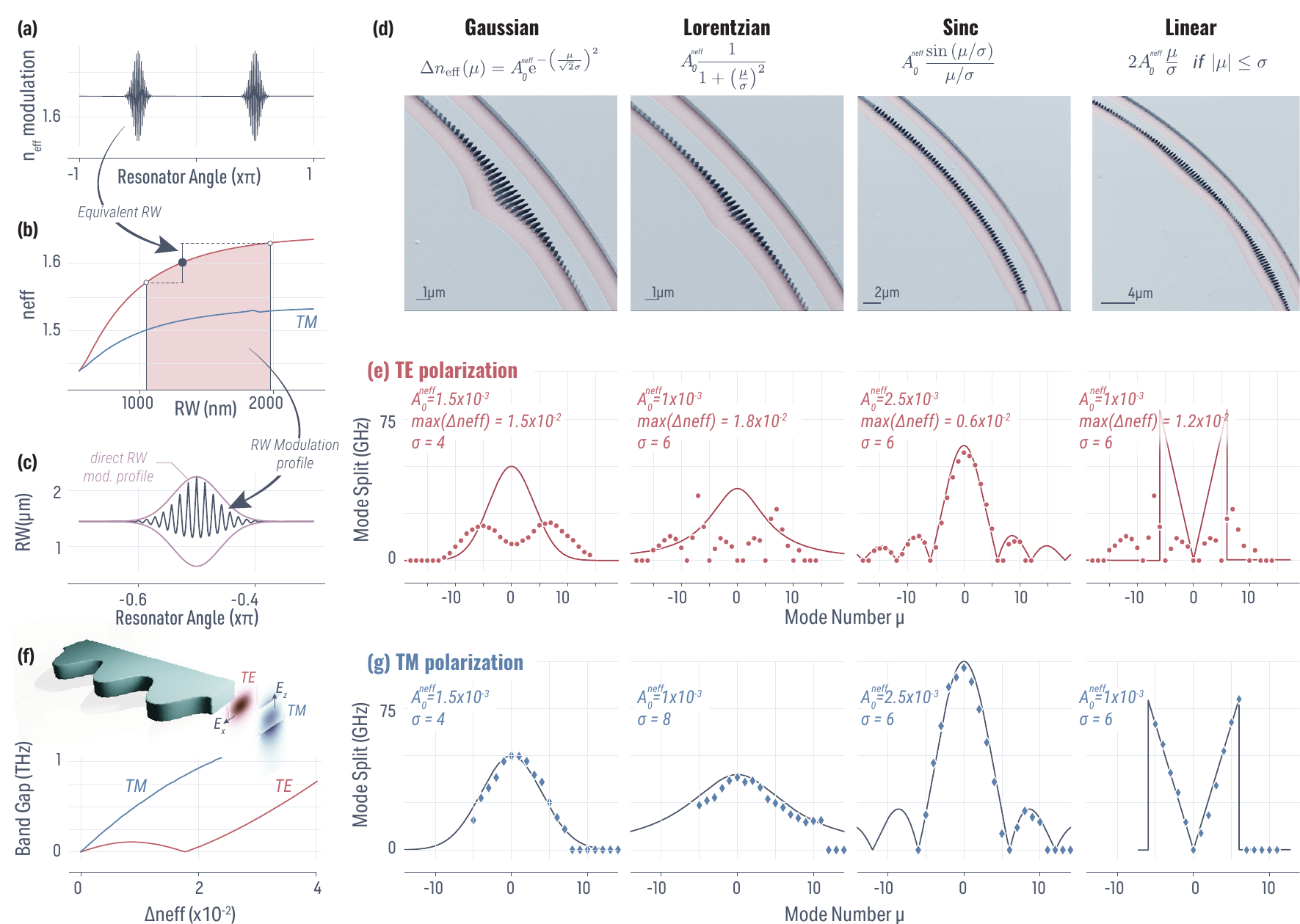}
    \caption{\label{fig:3} %
    \textbf{Improved Fourier Synthesis dispersion engineering considering the effective index for transverse electric and magnetic fundamental mode. } %
    \textbf{(a)} Instead of focusing on direct $RW$ modulation, we focus on the effective index modulation, for which we apply the DFT and obtain the total $n_\mathrm{eff}$ modulation along the azimuthal angle of the resonator. %
    \textbf{(b)} We can then map the $RW$ modulation from this $n_\mathrm{eff}$ profile using FEM simulation for either TE or TM polarization. %
    \textbf{(c)} Zoom-in of the resulting $RW$ modulation from this $n_\mathrm{eff}$ Fourier synthesis method. The direct $RW$ Fourier synthesis envelope is highlighted with a solid red line. We note that the minimal $RW$ is much larger with this method, preventing the two spurious effect previously presented. The modulation `teeth' are also sharper. %
    \textbf{(d)} Four different modal envelopes have been implemented: a Gaussian, Lorentzian, sinc, and linear. Each of them has been fabricated following the $n_\mathrm{eff}$ Fourier synthesis method. \textbf{(e)} Experimental result of the mode splitting (\textit{i.e.} CW/CCW coupling) for each resonance and each envelope for the TE polarization. Only the sinc envelope follows the designed modal envelope, which we explain in the main text. %
    \textbf{(f)} Fundamental consideration of the polarization needs to be accounted for, where the TE modes show discontinuity of the field at the modulated interface, while TM ones are not impacted. This results in a difference of band-gap behavior where the TE mode needs to account for a corrective term in the CW/CCW coupling~\cite{LeeOpt.Lett.2019}, resulting in a band gap closing. The TM mode, however, has a continuous increase of the band gap with increasing $n_\mathrm{eff}$ gradient. % 
    \textbf{(g)} Measuring the TM polarization, where each designed modal envelope is reproduced by the mode-splitting experiment, as expected from the TE/TM band gap behavior difference. In \textbf{(e)} and \textbf{(g)}, the one standard deviation uncertainty in splitting values is smaller than the data point size. % 
    }
\end{figure*}

\begin{figure*}[t]
    \centering
    \includegraphics[width=\textwidth]{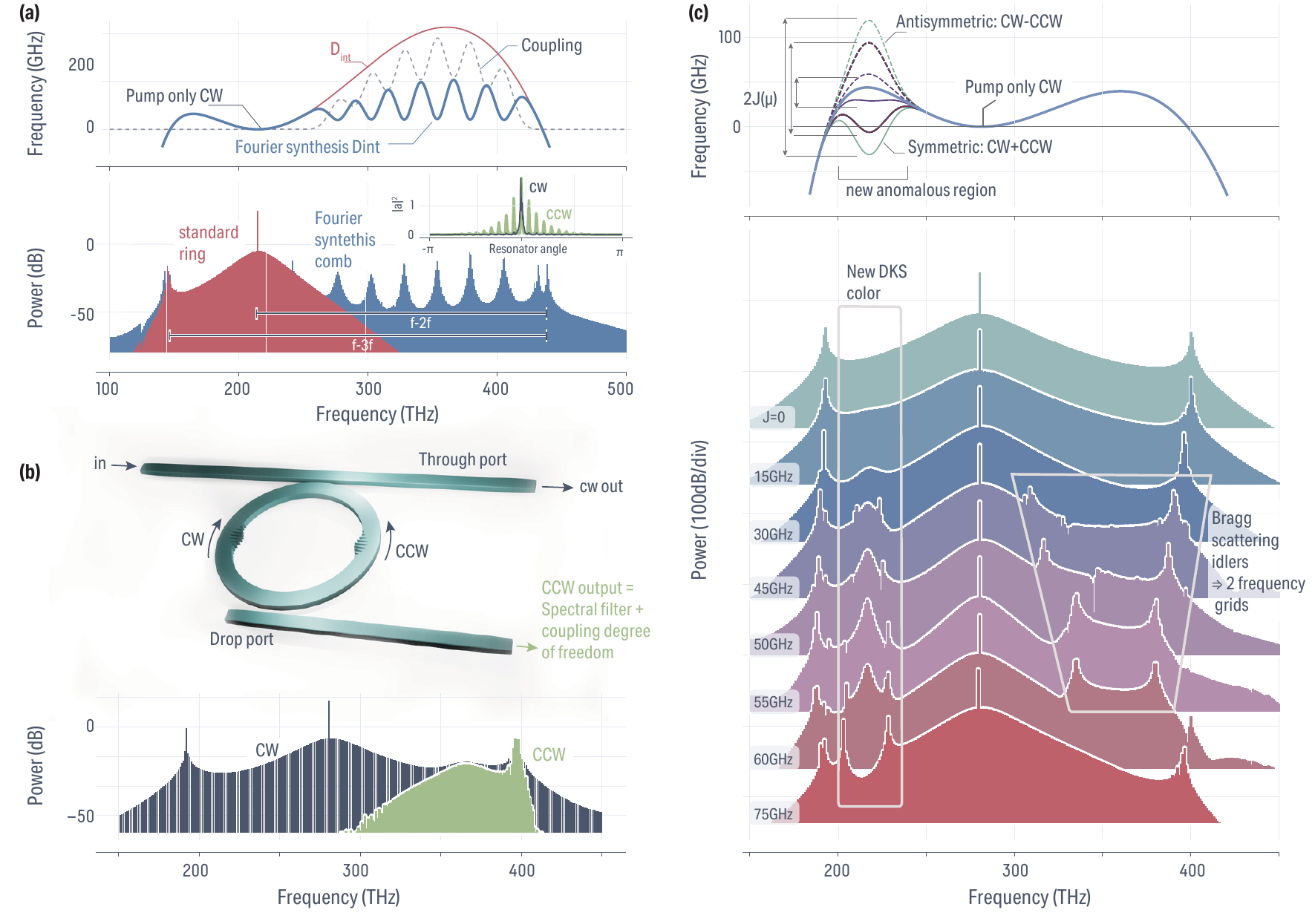}
    \caption{\label{fig:4} %xz
    \textbf{Numerical investigation of frequency comb tailoring from Fourier synthesis dispersion engineering.} %
    \textbf{(a)} Through $RW$ and $H$ dispersion engineering design, it is possible to obtain a harmonic relationship between the pump and $D_\mathrm{int}=0$ position (i.e., potential DW position). Yet, as presented in the first section, the soliton does not have sufficient power at this spectral location, resulting in essentially no DW at high frequency (red). Yet, through Fourier synthesis, one could lower the $D_\mathrm{int}$ barrier so that enough power can be transferred to the DW. The design of the Fourier synthesis $D_\mathrm{int}$ is trivial as the only criterion is to lower it enough to be close to zero. Here we use multiple Gaussian modal envelopes resulting in a modulated $D_\mathrm{int}$ (blue). The resulting spectrum remains the same around the pump with additional new soliton colors thanks to the anomalous dispersion pocket created by the Fourier synthesis. This allows having enough energy at the $D_\mathrm{int}$ zero-crossing to generate a significant DW which is harmonic with the pump. The inset represents the azimuthal profile of the soliton, with the CW component remaining close to a regular DKS and the CCW being a train of pulses. %
    \textbf{(b)} The CCW component being a soliton allows for frequency filtering. Complex interposer systems have been proposed to process a DKS microcomb. Here we propose that our Fourier synthesis system's intrinsic CW/CCW coupling allows for filtering the different comb spectral regions. For instance, by creating a CW/CCW coupling at high frequency (short wavelength), one could simply outcouple the CW light only to obtain these comb teeth without the main pump. %
    \textbf{(c)} We investigate the property of the anomalous dispersion pockets that are created by our Fourier synthesis. We use the same Gaussian modal profile with different amplitude, resulting in an anomalous region minimum that is at different frequency detuning from $D_\mathrm{int}=0$. Through coupled LLE simulation, we notice that a new soliton color is created in this anomalous pocket. When this pocket goes below $D_\mathrm{int}=0$, new DWs are created on the other side of the pump. This behavior is similar to~\cite{MoilleNat.Commun.2021}, and highlights that the soliton in the anomalous pocket from Fourier synthesis lives at a fixed carrier envelope offset shift from the main soliton, resulting in a nonlinear coupling in the phase space that creates new idlers from four-wave mixing Bragg scattering. %
    }
\end{figure*}

\vspace{1ex}\noindent\textbf{Effective index approach: improved Fourier synthesis and polarization considerations}

We have demonstrated that direct Fourier synthesis for predictable dispersion engineering using a simple $RW$ modulation approach is inherently flawed when modulation amplitudes are sufficiently large. However, given that PhC band structures are intrinsically related to the modal $n_\mathrm{eff}$, in this section we consider whether we can produce a better model for dispersion engineering with DFT design. 

The modal effective refractive index is not linear with the $RW$, given that cut-off occurs in our asymmetrically air-clad system under a small enough $RW$  and will grow asymptotically toward the silicon nitride index with large $RW$ . Thus it is more suitable to apply the inverse DFT on the effective index modulation rather than the RW modulation [\cref{fig:3}(a)]. The mapping between effective index $n_\mathrm{eff}$ and $RW$ is easily achieved through finite element method simulations. Of course, this function will also depend on thickness and ring radius, yet we assume they are fixed throughout this study. From the inverse DFT $n_\mathrm{eff}$ profile along the resonator angle, it is then possible to map the actual $RW$ modulation that we have to fabricate using this simulated calibration of $n_\mathrm{eff}$ [\cref{fig:3}(b)]. It is instructive to notice the difference between the modulation profile obtained by the previous direct $RW$ modulation method and the one described in this section [\cref{fig:3}(c)]. In the former case, the maximum $RW$ modulation is close to the nominal one, creating a local $RW$ close to zero, which is responsible for the cavity-like and higher-order mode coupling presented earlier. It can be understood that the local variation of $n_\mathrm{eff}$ is much stronger than expected, creating a local cut-off that leads to strong Bragg reflection at that point. However, in the $n_\mathrm{eff}$ modulation case, the $RW$ modulation accounts for the nonlinear dependence of $n_\mathrm{eff}$ with the ring width, in particular, the narrower it becomes. Therefore, the ring does not present the same bottle-neck effect, which reduces the two spurious effects partially responsible for the mode splitting not following the intended design. In addition, the ``teeth'' of the PhC are much sharper than in the simple sine modulation case, which we believe also reduces the PhC cavity mode within the Bragg grating. This also suggests that the upper modulation limit is not the unperturbed ring width, but is rather limited by either the disk $n_\mathrm{eff}$ (\textit{i.e.} $RW\rightarrow RR$) or the cut-off $RW$. This maximum modulation can be tailored with the nominal $RW$, but a trade-off must be found to allow for operating in the correct dispersion regime of the resonator for a given application (e.g., weak anomalous dispersion for a bright, broadband soliton microcomb).

Based on this new inverse DFT process, we proceed to fabricate four types of modal coupling designs (Fig.~\ref{fig:3}(d)) made around the same microring parameters presented in the previous section. We implement the same Gaussian profile, as well as a Lorentzian one defined by $n_\mathrm{eff} ^{(\mathrm{Lor})}(\mu) = A_0^\mathrm{neff} \frac{1}{1+(\mu/\sigma)^2}$, a sinc pattern defined by $n_\mathrm{eff} ^{(\mathrm{sinc})}(\mu) = A_0^\mathrm{neff} \frac{\mathrm{sin}\left(\mu/\sigma\right)}{\mu/\sigma}$, and a linear profile $n_\mathrm{eff} ^{(\mathrm{lin})}(\mu) = 2A_0^\mathrm{neff} \frac{\mu}{\sigma}$ with $\mu \in [-\sigma, \sigma]$ and no coupling outside this range for this profile [\cref{fig:3}(e)]. Measuring the TE\textsubscript{0} modes, we retrieve the mode splitting for each resonance, where once again, experimental data and the modal coupling design do not match [\cref{fig:3}(e)], except for the sinc profile. 

To understand the origin of this discrepancy, it is important to recall some basic boundary conditions put forth in Maxwell's equation, that is, that the electric field is discontinuous at the orthogonal interfaces of the polarization. While in this work we have used a simple perturbative expression for the coupling term between the CW and CCW modes (albeit while taking into account the waveguide effective modal index dependence on width), more generally, the coupling term between the CW and CCW modes can be semi-analytically determined by assuming a one-dimensional Bragg grating with two coupled counterpropagating waves. The transverse magnetic (TM) polarization -- with the dominant electric field component along the thickness direction of the ring -- does not experience a discontinuity at the modulated interface, which is also true for the magnetic field in this case along the radial direction [\cref{fig:3}(f)]. Therefore, the coefficient between CW/CCW coupling can be expressed as $J \propto \int_r\frac{1}{\varepsilon} H(r)H(r)^*\mathrm{d} r$ with $H(r)$ the magnetic mode profile along the radial direction of the ring. However, the TE mode experiences a discontinuity along the modulated ring wall. Thus, one needs to account for a correction term in the coupling such that $J \propto \int \varepsilon \left(E(r)E(r)^*-\partial_r E(r)\partial_r E(r)^*\right)\mathrm{d} r$~\cite{LeeOpt.Lett.2019}. This translates into different behaviors of the band-gap (\textit{i.e.} splitting amplitude) between the two polarizations, where the TM mode exhibits a continuously increasing band gap while the TE mode undergoes a band gap closing with the increase of the refractive index contrast of the PhC (\cref{fig:3}(f), band gap obtained from FEM simulation). This difference in behavior can help explain the mismatch between design and experimental data in the TE polarization. Interestingly, the sinc profile effectively is a single targeted modulation $M_0$ that is truncated to match the sinc modal width. Therefore, much less effect due to the polarization will occur and the behavior is closer to the single mode splitting case, explaining the good agreement with the designed modal mode splitting profile. In addition, the maximum $n_\mathrm{eff}$ modulation for a given $\sigma$ and $A_0$ is much smaller in the case of the sinc profile, which makes the system behave in the linear regime of band gap opening with modulation amplitude (e.g., as shown in Fig.~\ref{fig:3}(f)). 

The above explanation suggests that the TM polarization should not suffer from such a theory-experiment disagreement for any profile. We measured the same microrings in the TM mode, where $M_0 = 392$ is resonant at 204.6~\unit{THz} [\cref{fig:3}(g)]. As expected from the absence of band gap closing due to the field continuity with the $RW$ modulation, the TM mode follows the intended modal coupling design for all of the profiles implemented. We note that the maximum obtained mode-splitting reached up to 100~\unit{GHz} (\textit{i.e.} coupling $J\simeq50$~\unit{GHz}), which is compatible with integrated dispersion mitigation for ultra-broad microcomb generation. Nevertheless, it is fair to assume that larger coupling strength can be easily implemented. The quality factor for both TE and TM is not particularly impacted in the spectral region where the coupling happens and remains at the $>5\times10^5$ level (see supplementary \cref{fig:SupMatFig1}). However, we believe that substantial losses will be introduced at shorter wavelengths. First, approaching the mode $2M_0$, the band folding will not occur between the two contra-propagative modes but instead with the $\Gamma$ direction in the reciprocal space of the PhC (\textit{i.e} close to the surface normal direction). Therefore, the light will be extracted almost vertically. This is relevant for optical angular momentum generation~\cite{LuArXiv2022,WangArXiv2022}, while for our frequency comb application, it will generate considerable loss given the strength of the PhC. Secondly, at short wavelengths, the $RW$ modulation will not be subwavelength anymore, and will essentially act as a scattering element which is likely to cause losses.\\%
\indent Going forward, we note that with accurate modeling of the TE coupling between CW and CCW modes and the experimental demonstration of the band gap closing - which has already been observed in a large modulated PhC ring~\cite{LuNat.Photon.2022} - along with accurate predictions of the effective modal refractive index, it may be possible to effectively develop Fourier synthesis dispersion engineering accurately in this polarization, though it may also be the case that full numerical solution of Maxwell's equations within a modulating microring might be needed for accurate prediction of the dispersion. This work is outside of the scope of this manuscript; instead, in the next section, we focus on how understanding how the demonstrated Fourier synthesis dispersion engineering technique can impact microresonator frequency comb generation.

\vspace{1ex}\noindent\textbf{Broadband microcomb dispersion engineering and Bragg multi-color solitons using the Fourier synthesis approach}. In this section, we numerically investigate the potential that the inverse DFT dispersion design could bring to ultra-broadband microcombs and the new kinds of microcombs that could be generated using this technique.

To accurately model the system, we use a set of linearly coupled Lugiato-Lefever equations (LLEs) that account for the CW and CCW interaction in the PhC ring. To account for the cross-phase modulation, we assume that the contra-propagative light will travel at twice the speed of light, given that the LLE is intrinsically in the moving frame of the light. Therefore, the cross-phase modulation can neglect the fast oscillations of the light and simply use the average of the contra-propagative mode power~\cite{FanOpt.Lett.2020}: 

\begin{align}
    \dot{a}_\mathrm{\uparrow, \downarrow}(\theta, t) =& \left(-\frac{\kappa}{2} + i\delta \omega \right) a_\mathrm{\uparrow, \downarrow} 
  + i \sum_\mu D_\mathrm{int}(\mu)\mathrm{e}^{-i\mu\theta}A_{\uparrow, \downarrow}(\mu, t) \nonumber\\
	- i & \sum_\mu J(\mu) \mathrm{e}^{-i\mu\theta}A_{\downarrow , \uparrow}(\mu, t)  \nonumber\\
	+ i&\gamma L \left(|a_{\uparrow, \downarrow} |^2 + 2\int_{-\pi}^{\pi} \frac{|a_ {\downarrow , \uparrow} |^2}{2\pi}\mathrm{d}\theta \right)a_ {\uparrow, \downarrow}  \nonumber\\
	+& \delta_{\uparrow} \sqrt{\kappa_\mathrm{ext} P_\mathrm{\uparrow}}
\end{align}

\noindent where $a_{\uparrow}$ and $a_{\downarrow}$ are the envelope field in the CW and CCW mode respectively, $\dot{a} = \partial a/\partial t$ the temporal derivative, $A_{\uparrow, \downarrow}(\mu, t) = \mathrm{FT}[a_{\uparrow, \downarrow}(\theta, t)]$, the Fourier transform of the field in the azimuthal direction, $J(\mu)$ is the coupling between CW/CCW mode $\mu$ arising from the PhC, which is mode dependant as previously designed and exhibits a minus sign because of the $\pi$ phase shift induced by the reflection, $\kappa/2\pi =755$~MHz and $\kappa_\mathrm{ext} = \frac{1}{2} \kappa$ are the total loss rate and the coupling loss rate respectively (corresponding to a coupling and intrinsic quality factor $Q_\mathrm{i} = Q_\mathrm{c}=7.50\times 10^5$ at 283~THz, following Refs.~\cite{MoilleOpt.Lett.OL2021,MoilleAPLPhotonics2022}), $\delta\omega$ is the pump detuning relative to the pumped mode $\mu=0$, $L = 2\pi\times2\times23$~{\textmu}m is the resonator circumference, $\gamma = 2.3$~W\textsuperscript{-1}.m\textsuperscript{-1} is the effective Kerr nonlinearity, $P_\mathrm{\uparrow}=150$~mW is the input pump power in the CW mode, and $\delta_\mathrm{\uparrow}$ is the Kronecker delta function. %
In the case of interest in this manuscript, we will only focus on Fourier synthesis outside of the pumping region to tailor the integrated dispersion. Thus, our model assumes that any azimuthal component $\mu$ presenting $J(|\mu|\gg 0)\neq 0$, namely where the presented CW/CCW engineered mode coupling occurs, is far from the pump mode. Therefore, at the pump,  $J(\mu = 0)=0$ and only one propagative mode is excited in the LLE model. As stated earlier in \cref{fig:1}(a), the current state-of-the-art for dispersion engineering is to design both ring width and thickness to achieve the desired dispersion regime and comb width, which ultimately leads to a trade-off between bandwidth and flatness. Using our Fourier synthesis technique, one could use a very asymmetric $D_\mathrm{int}$ where the pump frequency and potential short dispersive wave position should be harmonic [\cref{fig:4}{a}] {corresponding to an oxide-clad device with ring width $RW=1125$~nm and thickness $H=770$~nm}. However, the $D_\mathrm{int}$ barrier being too large and steep, no power is transferred through Cerenkov radiation to create an appreciable DW. Using a simple Gaussian spectral coupling profile, whicsh can be added up around different modes to effectively decrease this barrier, it is not only possible to allow for sufficient energy transfer to create an appreciable DW [\cref{fig:4}(a)], but also reach a level of flatness through the different pockets of anomalous dispersion that have been created through Fourier synthesis dispersion engineering. The azimuthal profile of the pulse remains similar to a DKS, while the CCW light, which only exists where the coupling is not zero, exhibits a pulse-train-like behavior. 

In addition, the concept that CCW light exists only at frequencies where coupling happens is of great interest for metrology application on-chip, where usually complex interposers have to be created to filter, guide, and interfere different portions of the microcomb spectrum~\cite{RaoLightSci.Appl.2021}. In the case of the Fourier synthesis microcomb, one could imagine simplifying such an interposer greatly by harnessing the CW/CCW nature of the intra-cavity field~[\cref{fig:4}(b)]. Using a second drop bus waveguide allows only extracting the short wavelength where the CW/CCW coupling occurs, simplifying the filtering and routing of the downstream interposer. 

Finally, we investigate the properties of the generated solitons in the presence of these new pockets of anomalous dispersion. As expected from previous studies~\cite{luo_multicolor_2016, MoilleOpt.Lett.OL2018}, additional anomalous spectral windows can facilitate the exchange of energy through Cherenkov radiation -- which is similar to the DW coupling to the soliton~\cite{akhmediev_cherenkov_1995, CherenkovPhys.Rev.A2017} -- as long as their relative dispersions overlap the same frequency grid. We investigate by applying the same Gaussian Fourier synthesis coupling to a regular integrated dispersion profile from a previously studied microring resonator~\cite{MoilleAPLPhotonics2022}, corresponding to $RW=1060$~nm and $H=440$~nm for the air-clad \ce{Si3N4} microring, and only the coupling amplitude $J$ is modified~[\cref{fig:4}(c)]. Once the coupling is strong enough that the integrated dispersion in this spectral region becomes close enough to zero, Cherenkov radiation is permitted, allowing for the soliton to exchange energy with this other color and ultimately leading to a multi-color soliton. Interestingly, the second color of the soliton is entirely driven by the PhC and the Bragg reflection nature of the ring, creating a multi-color Bragg soliton~\cite{EggletonJ.Opt.Soc.Am.BJOSAB1999, WangLightSciAppl2020a}. Increasing the coupling to reach a negative value of the integrated dispersion in the CW/CCW coupling region still allows for a multi-color soliton to exist. However, it is interesting to notice a new set of DWs on the other spectral side of the pump. These DWs -- or idlers -- are signs of two solitons traveling at the same speed (\textit{i.e.} repetition rate) through cross-phase modulation coupling~\cite{WangOptica2017}, yet offset from one another, which has been observed in photonic molecules~\cite{HelgasonNat.Photonics2021, TikanNat.Phys.2021a} and multi-pump DKS~\cite{MoilleNat.Commun.2021} where such frequency interleaving might be harnessed for metrology applications. 

\vspace{1ex}\noindent\textbf{\large Discussion}. 

\noindent We have demonstrated a thorough approach for Fourier synthesis dispersion engineering of microring resonators. By summing the modulation at different amplitudes targeting several azimuthal modes, the azimuthal ring width variation can be obtained by an inverse discrete Fourier transform. We note that photonic molecules ~\cite{HelgasonarXiv2022, HelgasonNat.Photonics2021} (\textit{i.e.,} created by the coupling of two ring resonators) provide another mechanism to significantly alter dispersion compared to single ring resonators. In contrast to photonic molecules, where the coupling between the two rings' modes is dependent on the physical gap between them and is not dispersive enough to tailor modal frequencies on a mode-by-mode basis, our approach allows for a complete arbitrary spectral envelope (\textit{i.e.} on-demand dispersion) by addressing modes individually. We further demonstrated that the simple ring width modulation approach is insufficient for the predictive design of avoided mode crossings between CW and CCW modes created by the photonic crystal. Instead, an effective refractive approach inverse DFT, which is then mapped onto the ring width modulation, provides a much better accuracy for the mode splitting design. In addition, we discuss the implication of the polarization of the modes, where the transverse electric modes are intrinsically more complex to Fourier synthesize than the transverse magnetic modes due to electric field discontinuity along the ring width modulation. We demonstrated that using TM polarization, one can accurately reproduce a coupling design experimentally over twenty modes, following any arbitrary spectral envelope.  Finally, we discuss through numerical simulations how such Fourier synthesis dispersion engineering can profoundly impact ultra-broadband microcombs and lead to interesting new physics with multi-color Bragg solitons. Although our work has focused on microcomb dispersion engineering using an effective refractive index profile that follows a sine modulation, it can be extended to other types of approach. First, any other periodic modulation profile, such as bead-like~\cite{HuangIEEEPhotonicsTechnol.Lett.2015}, rectangular~\cite{deGoedeOpt.Express2021, BrunettiJ.Opt.2020}, or spike-like~\cite{ChenarXiv2022} profiles studied previously can be used in a Bloch expansion and hence are amenable to our Fourier synthesis approach. Moreover, in our photonic crystal rings, the Brillouin zone edge remains below the light cone, yet coupling to the light cone can happen with half the modulation period and can find application in the ejection of orbital angular momentum beams~\cite{MiaoScience2016} and the creation of optical vortex-generating microcombs~\cite{LiuarXiv2022,ChenarXiv2022}. Our work can, for example, allow for multiplexing multiple OAM states for each individual comb line of a microcomb.

%  .oooooo..o                             ooo        ooooo               .   
% d8P'    `Y8                             `88.       .888'             .o8   
% Y88bo.      oooo  oooo  oo.ooooo.        888b     d'888   .oooo.   .o888oo 
%  `"Y8888o.  `888  `888   888' `88b       8 Y88. .P  888  `P  )88b    888   
%      `"Y88b  888   888   888   888       8  `888'   888   .oP"888    888   
% oo     .d8P  888   888   888   888       8    Y     888  d8(  888    888 . 
% 8""88888P'   `V88V"V8P'  888bod8P'      o8o        o888o `Y888""8o   "888" 
%                          888                                               
%                         o888o                                              

\vspace{2ex}\noindent\textbf{\large Methods}

\noindent\textbf{Additional data on mode splitting measurements}

\begin{figure}[h]
    \centering
    \includegraphics[width = \columnwidth]{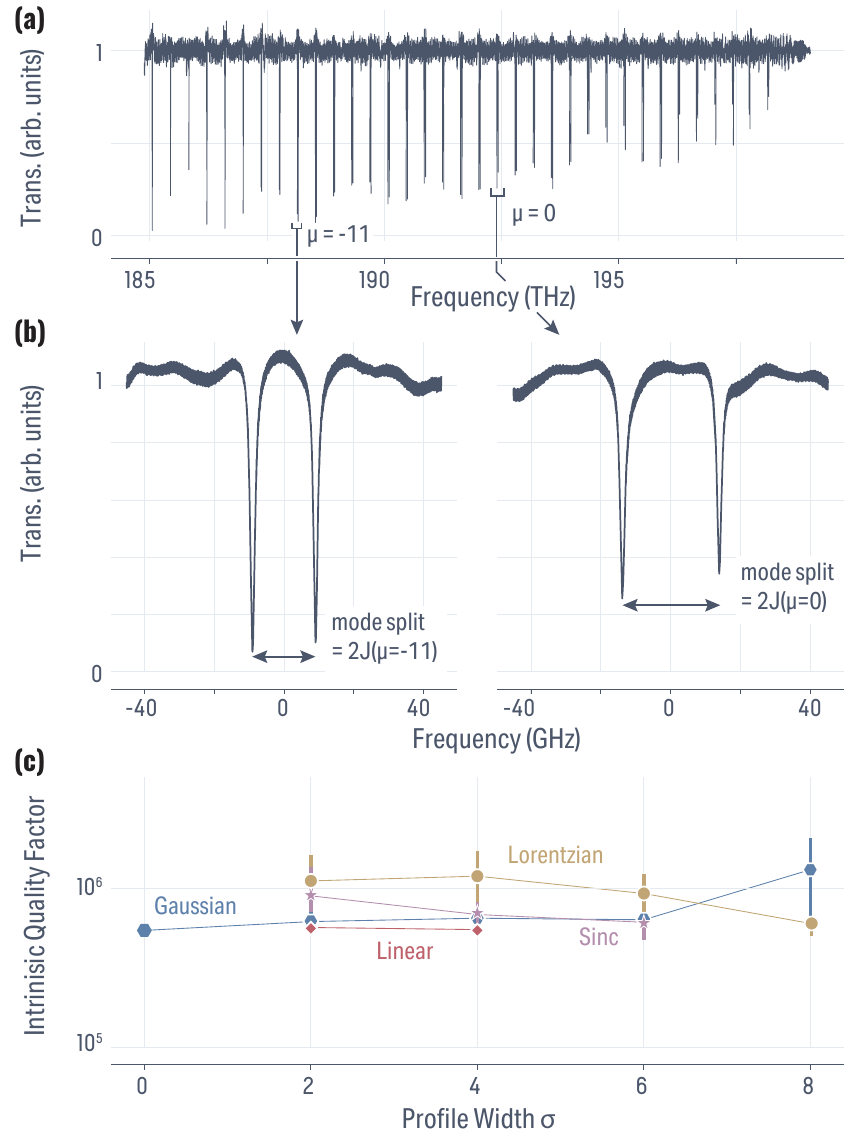}
    \caption{\label{fig:SupMatFig1} %
    \textbf{Linear properpy of Fourier synthesis photonic crystal ring resonator. }%
    \textbf{(a)} Representative transmission spectrum of TE mode a measured photonic crystal ring resonator. We normalize the mode numbers $\mu$ from the center of the modal coupling envelope. \textbf{(b)} Zoom-in onto two set of mode at $\mu=0$ and $\mu=-11$, from which we extract the coupling from the mode splitting. \textbf{(c)} Intrinsic quality factor extracted for each modal envelope presented in \cref{fig:3}, where the data point represents the average value and the error bar is the standard deviation resulting from variation in $Q$ across the mode splitting envelope (in some cases, the error bar is smaller than the data point size).  The overall $Q$ remains the same and one could conclude that the $n_\mathrm{eff}$ Fourier synthesis does not impact the intrinsic optical loss property of the ring. The error bar represent a standard variation obtain from the distribution of the Q over the resonance that are probed betwen 1510~nm and 1630~nm. When not displayed, the standard deviation is smaller than the marker.
    }
\end{figure}

\noindent The mode-split and contra-propagative coupling strengths we present throughout the manuscript are extracted from linear measurements such as the one presented in~\cref{fig:SupMatFig1}(a). In the case of the TE mode spectroscopy ranging from about 185~THz and 198~THz, the continuously tunable laser (CTL) is calibrated by simultaneously probing a tri-gas cell. Using the NIST calibration of \ce{H12CN}, \textsuperscript{13}\ce{CO} and \textsuperscript{13}\ce{CO} absorption lines we are able to accurately retrieve the wavelength of the transmission scan. We define the normalized mode number $\mu$ relative to the centered mode-splitting, which is retrieved with a single-mode splitting device (\textit{i.e.} sine modulation of the ring width), and each mode is then indexed from this mode, allowing to retrieve the coupling from the mode splitting [\cref{fig:SupMatFig1}(b)]. From this linear spectroscopy data, we also extract the intrinsic and the coupling quality factors through a simple coupled mode-theory model~\cite{BorselliOpt.ExpressOE2005}. In \cref{fig:SupMatFig1}(c), we present the average intrinsic quality factor for the four modal coupling envelopes presented in \cref{fig:3} for the TE polarization, with the nominal spectral size $\sigma$ for each envelope, with the error bar representing the standard variation of Q over the different modes measured. The intrinsic quality factor does not seem to be impacted and remains close to a million with little variation over the different modes, although the modulation of the ring width is extremely local (see the following section). The excess radiation losses induced by the photonic crystal modulation can therefore be concluded to be minimal in our system.

\noindent\textbf{Fabrication details}

\begin{figure}[t]
    \includegraphics[width = \columnwidth]{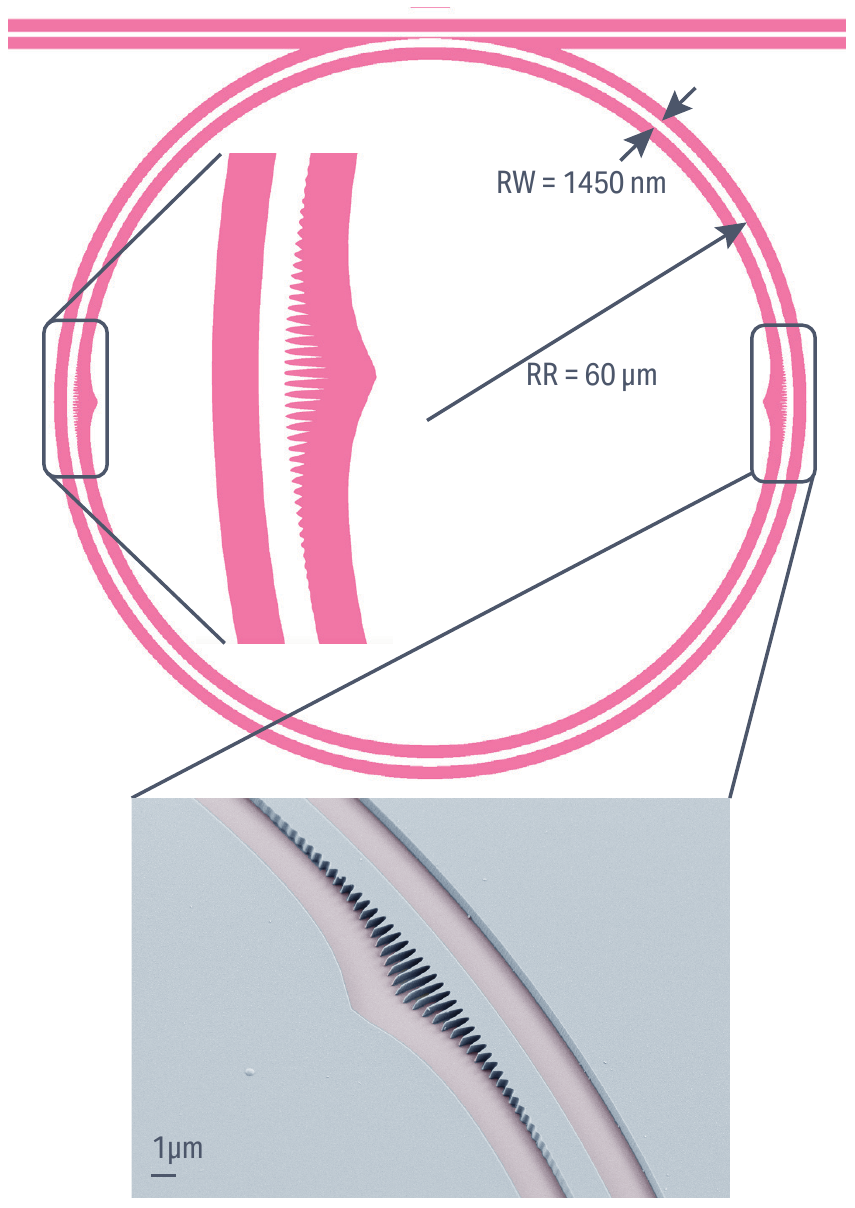}
    \caption{\label{fig:SupMatFig2} %
    \textbf{Fabrication and patterning. } %
     Fabricated layout for the case of a Gaussian envelope. The left inset represents a zoom-in of the layout where the localized modulation resulting from the $n_\mathrm{eff}$ Fourier synthesis occurs. The bottom inset is an SEM image of the fabricated ring where the modulation `teeth' are fully resolved. %
    }
\end{figure}

The microresonators are made of a 440~nm thick stoichiometric \ce{Si3N4} grown via low pressure chemical vapor deposition with a 7:1 ammonia/dichlorosilane gas ratio, on top of thermal \ce{SiO2}. The layout presented in~\cref{fig:SupMatFig2} is patterned using electron-beam lithography. The ring and waveguide are then etched using a reactive ion etching method. The facets are clad with \ce{SiO2} while the rings are still air-clad thanks to a lift-off process of the \ce{SiO2}. The chips are then diced and polished for testing.

\vspace{1ex}
\noindent \textbf{\large Data availability} \\
The data that supports the plots within this paper and other findings of this study are available from the corresponding authors upon reasonable request. 

\vspace{1ex}
\noindent \textbf{\large Code availability} \\
The LLE code to study the microcomb dynamics is a modified version of \textit{pyLLE}~\cite{MoilleJ.RES.NATL.INST.STAN.2019} which can be made available upon reasonable request.

%  oooooooooo.   o8o   .o8       oooo   o8o            
% `888'   `Y8b  `"'  "888       `888   `"'            
%  888     888 oooo   888oooo.   888  oooo   .ooooo.  
%  888oooo888' `888   d88' `88b  888  `888  d88' `88b 
%  888    `88b  888   888   888  888   888  888   888 
%  888    .88P  888   888   888  888   888  888   888 
% o888bood8P'  o888o  `Y8bod8P' o888o o888o `Y8bod8P' 

% \bibliography{Biblio}
% \bibliographystyle{naturemag}

% /*ooo        ooooo  o8o                     
% `88.       .888'  `"'                     
%  888b     d'888  oooo   .oooo.o  .ooooo.  
%  8 Y88. .P  888  `888  d88(  "8 d88' `"Y8 
%  8  `888'   888   888  `"Y88b.  888       
%  8    Y     888   888  o.  )88b 888   .o8 
% o8o        o888o o888o 8""888P' `Y8bod8P' 

\vspace{1ex}
\noindent \textbf{\large Acknowledgements} \\
The authors thank Sashank Shridar and Yi Sun for their valuable input. The authors acknowledge funding from the DARPA APHI and NIST-on-a-chip programs.
While preparing the manuscript, we have been made aware of Ref.\cite{LucasArXiv2022} that has some similarities to our approach. Our work goes beyond the regime of relatively small mode splittings to theoretically and experimentally treat the $\approx$100~GHz shifts needed for broadband dispersion engineering and multi-color soliton formation.

\vspace{1ex}
\noindent \textbf{\large Author contributions}\\
G.M. led the project, designed the ring resonators, conducted the experiments, and helped develop the theoretical framework. X.L and J.S. helped with the design, D.W fabricated the devices. G.M. and K.S. wrote the manuscript with input from all authors. All the authors contributed and discussed the content of this manuscript.

\vspace{1ex}
\noindent \textbf{\large Competing Interests}\\
The authors declare no competing interests.

\end{document}